\documentclass[article]{jss}

\usepackage{amsbsy} 
\usepackage{amsmath}
\usepackage{amsfonts}
\usepackage{thumbpdf}
\usepackage{subfigure}
\usepackage{float}
\usepackage{graphicx}
\usepackage{epstopdf}
\usepackage[latin1]{inputenc}

\author{Angelo  Mazza   \\ University of Catania    \And 
        Antonio Punzo   \\ University of Catania    
        }
\title{
\pkg{DBKGrad}: An \proglang{R} Package for Mortality Rates Graduation by Fixed and Adaptive\\ Discrete Beta Kernel Techniques
}

\Plainauthor{Angelo Mazza, Antonio Punzo} 
\Plaintitle{dbkGrad: An R Package for Graduation by Discrete Beta Kernel Techniques} 
\Shorttitle{\pkg{DBKGrad}: An \proglang{R} Package for Graduation by Discrete Beta Kernel Techniques} 

\Abstract{
Kernel smoothing represents a useful approach in the graduation of mortality rates.
Though there exist several options for performing kernel smoothing in statistical software packages, there have been very few contributions to date that have focused on applications of these techniques in the graduation context.
Also, although it has been shown that the use of a variable or adaptive smoothing parameter, based on the further information provided by the exposed to the risk of death, provides additional benefits, specific computational tools for this approach are essentially absent.
Furthermore, little attention has been given to providing methods in available software for any kind of subsequent analysis with respect to the graduated mortality rates.

To facilitate analyses in the field, the \proglang{R} package \pkg{DBKGrad} is introduced.
Among the available kernel approaches, it considers a recent discrete beta kernel estimator, in both its fixed and adaptive variants.
In this approach, boundary bias is automatically reduced and age is pragmatically considered as a discrete variable.
The bandwidth, fixed or adaptive, is allowed to be manually given by the user or selected by cross-validation.
Pointwise confidence intervals, for each considered age, are also provided.
An application to mortality rates from the Sicily Region (Italy) for the year 2008 is also presented to exemplify the use of the package.
}

\Address{
  Antonio Punzo\\
  Dipartimento di Economia e Impresa\\
  Universit{\`a} di Catania\\
  95129 Catania, Italia\\
  Telephone: +39/095/7537640\\
  Fax: +39/095/7537610\\
  E-mail: \email{antonio.punzo@unict.it}\\
  URL: \url{http://www.economia.unict.it/punzo}
}

\begin{document}


\section[Introduction]{Introduction}
\label{sec:Introduction}

Mortality rates are age-specific indicators commonly used in demography.
Historically, they are also widely adopted by actuaries, in the form of mortality tables, to calculate life insurance premiums, annuities, reserves, and so on.
Producing these tables from a suitable set of crude (or raw) mortality rates is called \textit{graduation}, and this subject has been extensively discussed in the actuarial literature (see, e.g., \citealt{Copa:nonp:1983} and \citealt{Habe:Rens:Gene:1996}).
To be specific, the $d_x$ deaths at age $x$ can be seen as arising from a population, initially exposed to the risk of death, of size $e_x$.
The situation is commonly summarized via the model $d_x\sim\text{Bin}\left(e_x,q_x\right)$, where $q_x$ represents the true, but unknown, mortality rate at age $x$.
The crude rate $\mathring{q}_x$ is the observed counterpart of $q_x$. 
Graduation is necessary because crude data usually presents abrupt changes, which do not agree to the dependence structure supposedly characterizing the true rates \citep{Lond:Grad:1985}. 
In fact, a common prior opinion about their form is that each true mortality rate is closely related to its neighbors.
This relationship is expressed by the belief that the true rates progress smoothly from one age to the next. 
So, the next logical step is to graduate the crude rates to produce smooth estimates, $\widehat{q}_x$, of the true rates. 
This is done by systematically revising the crude rates in order to remove any random fluctuations. 
Nonparametric models are the natural choice if the aim is to reflect this belief.
Furthermore, a nonparametric approach can be used to choose the simplest suitable parametric model, to provide a diagnostic check of a parametric model, or to simply explore the data (see \citealt[][Section~1.1,]{Hard:Appl:1992} for a detailed discussion on the chief motivations that imply their use, and \citealt{DMSa:acom:2006} for an exhaustive comparison of nonparametric methods in the graduation of mortality rates).

Due to its conceptual simplicity and practical and theoretical properties, kernel smoothing is one of the most popular statistical methods for nonparametric graduation. 
Among the various alternatives existing in literature (see \citealt{Copa:nonp:1983}, \citealt{Gavi:Habe:Verr:Movi:1993,GHVe:onth:1994,GHVe:grad:1995} and \citealt{Peri:Kost:Anev:2005}), the attention is here focused on the discrete beta kernel estimator proposed by \citet{Mazz:Punz:Disc:2011}.
Roughly speaking, the genesis of this model starts with the consideration that, although age $X$ is in principle a continuous variable, it is typically truncated in some way, such as age at last birthday, so that it takes values on the discrete set $\mathcal{X}=\left\{0,1,\ldots,\omega\right\}$, $\omega$ being the highest age of interest.
Discretization of age, from a pragmatical and practical point of view, could also come handy to actuaries that have to produce ``discrete'' graduated mortality tables starting from the observed counterparts. 
In the fixed bandwidth estimator proposed in \citet{Mazz:Punz:Disc:2011}, the discrete beta probability mass functions of \citet{Punz:Zini:Appr:2012}, parameterized according to \citet[][see also \citealp{Bagn:Punz:Fine:2013}]{Punz:disc:2010}, are considered as kernel functions in order to overcome the problem of boundary bias, commonly arising from the use of symmetric kernels \citep[see][]{Chen:beta:2000}.
The support $\mathcal{X}$ of the discrete beta, in fact, matches the age range and this, when smoothing is made near the boundaries, allows avoiding allocation of weight outside the support (for example negative or unrealistically high ages). 
Variants of the fixed bandwidth discrete beta kernel estimator, which allow the bandwidth to vary at each age according to the reliability of the data, also exist; in \citet{Mazz:Punz:Grad:2012}, the reliability is expressed by the $e_x$, while in \citet{Mazz:Punz:Usin:2013} this reliability is measured via the reciprocal of the variation coefficient (VC), with the VC being function of both the amount of exposure and the observed mortality rate.

In this paper we present the \proglang{R} \citep{R} package \pkg{DBKGrad}, available from CRAN (\url{http://CRAN.R-project.org/}), which offers all the features described above and adds some related functionalities.
Although \proglang{R} is well-provided with kernel smoothing techniques \citep[see, e.g.,][]{Hayfield:Racine:2008:JSSOBK:v27i05}, it does not offer neither discrete beta kernel smoothing, nor application of kernel smoothing techniques in graduation of mortality data. 
Note that nonparametric smoothing techniques, of the kind found in \pkg{DBKGrad}, are commonly used and often cited exploratory statistical tools; as evidence, consider the number of times in which classical statistical studies use the functions \code{density} and \code{ksmooth}, both in the \pkg{stats} package, for kernel smoothing estimation of a density or regression function.

The paper is organized as follows.
Section~\ref{sec:Discrete beta kernel graduation} retraces the fixed discrete beta kernel estimator.
Its adaptive variants are recalled in Section~\ref{sec:Making the bandwidth adaptive} while some cross-validation approaches for the selection of both the fixed and the adaptive bandwidth is discussed in Section~\ref{sec:CV}.
Further related aspects, such as the adoption of a preliminary logit transformation of the rates and the computation of the pointwise confidence intervals, are given in Section~\ref{sec:kernel}.   
The relevance of the \pkg{DBKGrad} package is shown, via a real data set, in Section~\ref{sec:Package DBKGrad in use}, and conclusions are finally given in Section~\ref{sec:conclusions}.

\section[Discrete beta kernel graduation]{Discrete beta kernel graduation}
\label{sec:Discrete beta kernel graduation}

Given the crude rates $\mathring{q}_y$, $y\in\mathcal{X}$, the Nadaraya-Watson kernel estimator of the true but unknown mortality rate $q_x$, at the evaluation age $x$, is
\begin{equation}
\widehat{q}_x=\sum_{y\in\mathcal{X}}\frac{k_h\left(y;m=x\right)}{\displaystyle\sum_{j\in\mathcal{X}}k_h\left(j;m=x\right)}\mathring{q}_y=\sum_{y\in\mathcal{X}}K_h\left(y;m=x\right)\mathring{q}_y,\quad x\in\mathcal{X},
\label{eq:NW kernel}
\end{equation}
where $k_h\left(\cdot;m\right)$ is the \textit{discrete kernel function} (hereafter simply named \textit{kernel}), $m\in\mathcal{X}$ is the single mode of the kernel, $h>0$ is the (fixed) \textit{bandwidth} (or \textit{smoothing parameter}) governing the bias-variance trade-off, and $K_h\left(\cdot;m\right)$ is the \textit{normalized kernel}.
Since we are treating age as being discrete, with equally spaced values, kernel graduation by means of \eqref{eq:NW kernel} is equivalent to moving (or local) weighted average graduation \citep{GHVe:grad:1995}.

\begin{figure}[!ht]
\centering
\subfigure[$h=0.4$\label{fig:beta_h0.4}]
{\includegraphics[width=0.325\textwidth]{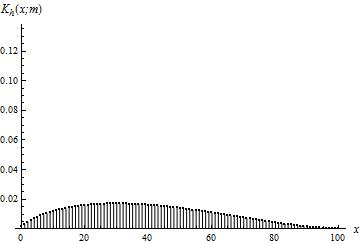}}
\subfigure[$h=0.04$\label{fig:beta_h0.04}]
{\includegraphics[width=0.325\textwidth]{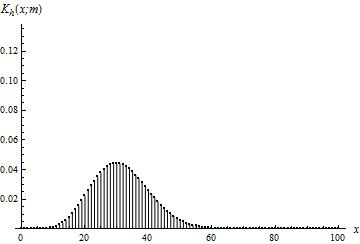}}
\subfigure[$h=0.004$\label{fig:beta_h0.004}]
{\includegraphics[width=0.325\textwidth]{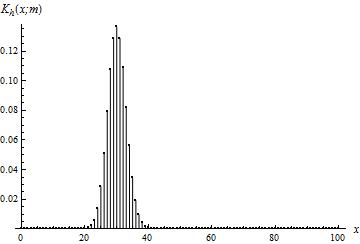}}
\caption{The effect of varying $h$ in the discrete beta probability mass function ($\omega=100$ and $m=30$).
\label{fig:beta h}}
\end{figure}

In \eqref{eq:NW kernel}, the discrete beta kernels \citep{Mazz:Punz:Disc:2011}
\begin{equation}
k_h\left(x;m\right)=\displaystyle\left(x+\frac{1}{2}\right)^{\frac{m+\frac{1}{2}}{h\left(\omega +1\right)}}\displaystyle\left(\omega +\frac{1}{2}-x\right)^{\frac{\omega +\frac{1}{2}-m}{h\left(\omega +1\right)}}
\label{eq:beta kernels}
\end{equation} 
are here adopted.
Their normalized version, 
\begin{displaymath}
K_h\left(x;m\right)=\frac{k_h\left(x;m\right)}{\displaystyle\sum_{y\in\mathcal{X}}k_h\left(y;m\right)},
\end{displaymath}
corresponds to the discrete beta probability mass function defined in \citet{Punz:Zini:Appr:2012} and parameterized, as in \citet{Punz:disc:2010}, according to the mode $m$ and another parameter $h$ that is closely related to the distribution variability.
In particular, for $h\rightarrow 0^+$, $K_h\left(x;m\right)$ tends to a Dirac delta function in $x=m$, while for $h\rightarrow\infty$, $K_h\left(x;m\right)$ tends to a discrete uniform distribution; \figurename~\ref{fig:beta h} shows the effect of varying $h$, maintaining constant $\omega$ and $m$.
Thus $h$ can be considered as the smoothing parameter of the estimator \eqref{eq:NW kernel}; indeed, as $h$ becomes smaller, the spurious fine structure becomes visible, while as $h$ gets larger, more details are obscured.

\begin{figure}[ht]
\centering
\subfigure[$m=0$]
{\includegraphics[width=0.325\textwidth]{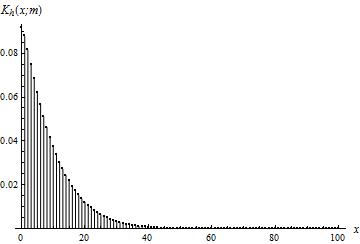}}
\subfigure[$m=5$]
{\includegraphics[width=0.325\textwidth]{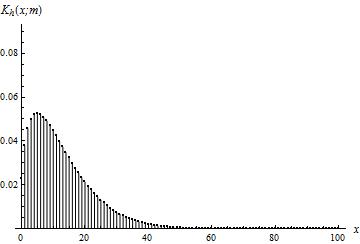}}
\subfigure[$m=10$]
{\includegraphics[width=0.325\textwidth]{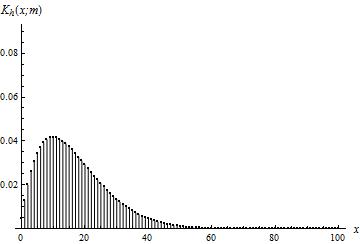}}
\subfigure[$m=20$]
{\includegraphics[width=0.325\textwidth]{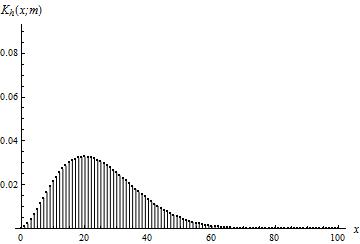}}
\subfigure[$m=35$]
{\includegraphics[width=0.325\textwidth]{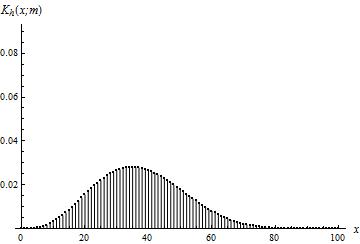}}
\subfigure[$m=50$]
{\includegraphics[width=0.325\textwidth]{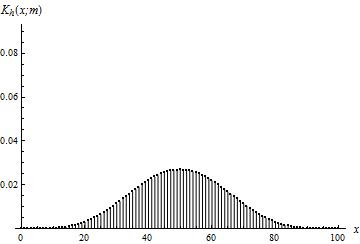}}
\caption{
The effect of varying $m$ in the discrete beta probability mass function ($\omega=100$ and $h=0.1$).
\label{fig:beta_m}
}
\end{figure}

Roughly speaking, discrete beta kernels possess two peculiar characteristics. 
Firstly, their shape, fixed $h$, automatically changes according to the value of $m$.
The graphical effect of varying $m$, keep fixed $h$ and $\omega$, is displayed in \figurename~\ref{fig:beta_m}.  
Secondly, the support of the kernels matches the age range $\mathcal{X}$ so that no weight is assigned outside the data support; this means that the order of magnitude of the bias does not increase near the boundaries.
Further details are reported in \citet{Mazz:Punz:Disc:2011}; see also \citet{Chen:beta:2000} to find out more on the properties of the discrete beta kernel estimator in its continuous counterpart.
The discrete beta kernel estimator is obtained with the specification \code{bandwidth="FX"} -- which represents the default -- in the \code{dbkGrad} function.

\section[Making the bandwidth adaptive]{Making the bandwidth adaptive}
\label{sec:Making the bandwidth adaptive}

Rather than restricting $h$ to a fixed value, a more flexible approach is to allow the bandwidth to vary according to the reliability of the data measured in a convenient way.
Thus, for ages in which the reliability is relatively larger, a low value for $h$ results in an estimate that more closely reflects the crude rates. 
For ages in which the reliability is smaller, such as at old ages, a higher value for $h$ allows the estimate of the true mortality rates to progress more smoothly; this means that at older ages we are calculating local averages over a greater number of observations.
This technique is often referred to as a variable or \textit{adaptive (bandwidth) kernel estimator} because it is characterized by an adaptive bandwidth $h_x\left(s\right)$ which depends on the reliability $l_x$ and is function of a further sensitive parameter $s$. 

Although the reliability $l_x$ can be inserted into the basic model \eqref{eq:NW kernel} in a number of ways \citep{GHVe:grad:1995}, here we adopt a natural formulation according to which 
\begin{equation}
h_x\left(s\right)=hl_x^s,\quad x\in\mathcal{X},
\label{eq:local bandwidth}
\end{equation}
where $h$ is the global bandwidth and $s\in\left[0,1\right]$.
Reliability decides the shape of the local factors, while $s$ is necessary to dampen the possible extreme variations in reliability that can arise between young and old ages.
Naturally, in the case $s=0$, we are ignoring the variation in reliability, which gives a fixed bandwidth estimator.

Using \eqref{eq:local bandwidth} we are calculating a different bandwidth for each age $x\in\mathcal{X}$, leading model \eqref{eq:NW kernel} to become
\begin{equation}
\widehat{q}_x=\sum_{y\in\mathcal{X}}\frac{k_{h_x}\left(y;m=x\right)}{\displaystyle\sum_{j\in\mathcal{X}}k_{h_x}\left(j;m=x\right)}\mathring{q}_y=\sum_{y\in\mathcal{X}}K_{h_x}\left(y;m=x\right)\mathring{q}_y,\quad x\in\mathcal{X},
\label{eq:adaptive NW kernel}
\end{equation}
where the notation $h_x$ is used to abbreviate $h_x\left(s\right)$.
Thus, for each evaluation age $x$, the $\omega+1$ discrete beta distributions $K_{h_x}\left(\cdot;m=x\right)$ vary for the placement of the mode as well as for their variability as measured by $h_x$.

In particular, \citet{Mazz:Punz:Grad:2012} consider the reliability a function only of the amount of exposure, according to the formulation   
\begin{equation}
l_x=\displaystyle\frac{f_x^{-1}}{\displaystyle\max_{y\in\mathcal{X}}\left\{f_y^{-1}\right\}},\quad x\in\mathcal{X},
\label{eq:ex local factor}
\end{equation}
where 
\begin{displaymath}
f_x=\displaystyle\frac{e_x}{\displaystyle\sum_{y\in\mathcal{X}}e_y}  
\end{displaymath}
is the empirical frequency of exposed to the risk of death at age $x$.
This alternative is allowed by the specification \code{bandwidth="EX"} in the \code{dbkGrad} function.

According to the model $d_x\sim\text{Bin}\left(e_x,\mathring{q}_x\right)$, where $\mathring{q}_x$ is the maximum likelihood estimate of $q_x$, a natural index of reliability is represented by the reciprocal of a relative measure of variability.
As relative measure of variability, \citet{Mazz:Punz:Usin:2013} adopt the variation coefficient (VC) which, in this context, can be computed as 
\begin{displaymath}
\text{VC}_x=\frac{\sqrt{e_x\mathring{q}_x\left(1-\mathring{q}_x\right)}}{e_x\mathring{q}_x},\quad x\in\mathcal{X}.
\end{displaymath}
It is inserted in \eqref{eq:local bandwidth} according to the formulation
\begin{equation}
l_x=\frac{\text{VC}_x}{\displaystyle\sum_{y\in\mathcal{X}}\text{VC}_y},\quad x\in\mathcal{X}.
\label{eq:VC local factor}
\end{equation}
In \eqref{eq:VC local factor}, $\text{VC}_x$ is normalized so that $l_x^s\in\left[0,1\right]$.
Note that reliability measured as in \eqref{eq:VC local factor} takes into account the amount of exposure $e_x$, but also the crude rate $\mathring{q}_x$.
The specification \code{bandwidth="VC"}, in the \code{dbkGrad} function, allows for this adaptive bandwidth variant.

\section{The Choice of $h$ and $s$}
\label{sec:CV}

As regard the fixed bandwidth estimator in \eqref{eq:NW kernel}, the choice of $h$ is important.
Although it is informative to choose the bandwidth by trial and error, it is also convenient to have an objective, risk-based method for selecting $h$.
The literature on data-driven methods for selecting the optimal value for $h$ is vast; however, cross-validation \citep{Ston:cros:1974} is without doubt the most commonly used and the simplest to understand. 
Cross-validation simultaneously fits and smooths the data by removing one data point at a time, estimating the value of the function at the missing point, and then comparing the estimate to the omitted, observed value. 
For a complete description of cross-validation in the context of graduation, see \citet{GHVe:grad:1995}. 
The cross-validation statistic to be minimized is
\begin{equation}
\text{CV}\left(h\right)=\sum_{x\in\mathcal{X}}r^2\left(\mathring{q}_x,\widehat{q}_x^{\left(-x\right)}\right)
\label{eq:CV statistic},
\end{equation}
where $r\left(\mathring{q}_x,\widehat{q}_x^{\left(-x\right)}\right)$ denotes the residual (at age $x$) and  
\begin{displaymath}
\widehat{q}_x^{\left(-x\right)}=\sum_{\substack{y\in\mathcal{X}\\
y\neq x}}\frac{K_h\left(y;m=x\right)}{\displaystyle\sum_{\substack{j\in\mathcal{X}\\
j\neq x}}K_h\left(j;m=x\right)}\mathring{q}_y
\end{displaymath}
is the estimated value at age $x$ computed by removing the crude rate $\mathring{q}_x$ at that age.
The bandwidth that minimizes $\text{CV}\left(h\right)$ is referred to as the cross-validation bandwidth.
As residuals, \citet{Mazz:Punz:Disc:2011,Mazz:Punz:Grad:2012} consider the classical residuals 
\begin{equation}
r\left(\mathring{q}_x,\widehat{q}_x^{\left(-x\right)}\right)=\widehat{q}_x^{\left(-x\right)}-\mathring{q}_x,
\label{eq:differences}
\end{equation}  
while \citet{Mazz:Punz:Usin:2013} adopt the proportional differences
\begin{equation}
r\left(\mathring{q}_x,\widehat{q}_x^{\left(-x\right)}\right)=\frac{\widehat{q}_x^{\left(-x\right)}}{\mathring{q}_x}-1,
\label{eq:proportional differences}
\end{equation}  
which is commonly used in the graduation literature because, since the high differences in mortality rates among ages, we want, in \eqref{eq:CV statistic}, the mean relative square error to be low \citep[see][]{Heli:Poll:Thea:1980}.
Cross-validation, with residuals \eqref{eq:differences}, is obtained with the specification \code{cvres="res"} while, with residuals \eqref{eq:proportional differences}, is obtained with the specification \code{cvres="propres"} (the default) in the \code{dbkGrad} function.

In the adaptive frame, in addition to the global bandwidth $h$, also the sensitivity parameter needs to be selected.
The natural choice consists in minimizing the bidimensional cross-validation statistic $\text{CV}\left(h,s\right)$ as a function of both $h$ and $s$ where in this case, $\widehat{q}_x^{\left(-x\right)}$ is naturally based on \eqref{eq:adaptive NW kernel}.
This is obtained via the specifications \code{cvh="TRUE"} and \code{cvs="TRUE"} in the \code{dbkGrad} function.
Nevertheless, in literature (see \citealp{GHVe:grad:1995} and \citealp{Mazz:Punz:Disc:2011,Mazz:Punz:Grad:2012,Mazz:Punz:Usin:2013}), $s$ is chosen subjectively and cross-validation is still used to select $h$ by minimizing the conditional cross-validation statistic $\text{CV}\left(h|s\right)$. 
This approach can be obtained by posing \code{cvh="TRUE"} and \code{cvs="FALSE"}, and by specifying a value for the argument \code{s} of the \code{dbkGrad} function. 
Note that, in the cross-validation routine, minimization is performed using the Levenberg-Marquardt algorithm \citep{More:TheL:1978} in the \pkg{minpack.lm} package \citep{minpack.lm}.   

\section{Further aspects}
\label{sec:kernel}

\subsection[The smoother matrix]{The smoother matrix}
\label{subsec:The smoother matrix}

Models \eqref{eq:NW kernel} and \eqref{eq:adaptive NW kernel} can be written, for notational and computational convenience, in the following compact (matricial) form
\begin{displaymath}
\boldsymbol{\widehat{q}}=\boldsymbol{K}\boldsymbol{\mathring{q}},	
\end{displaymath}
where $\boldsymbol{\mathring{q}}$ and $\boldsymbol{\widehat{q}}$ are the $\left(\omega+1\right)$-dimensional vectors of crude and graduated mortality rates, respectively, while $\boldsymbol{K}$ is the so-called $\left(\omega+1\right)\times\left(\omega+1\right)$ \textit{smoother} (or \textit{hat}) matrix -- depending on the bandwidth $h$ and eventually also on the sensitivity parameter $s$ -- in which the $i$-th row contains the $\omega+1$ weights allocated to $\mathring{q}_x$, $x\in\mathcal{X}$, in order to obtain $\widehat{q}_{i-1}$.
The smoother matrix is one of the values, named \code{kernels}, returned by the \code{dbkGrad} function.

\subsection[Transforming mortality rates]{Transforming mortality rates}
\label{subsec:Transforming mortality rates}

Before applying any model, it is always worth considering a transformation of the data into a more tractable form, that better reflects the strengths of the model or that more clearly reveals the structure of the data.
In parametric graduation, for example, it may be easier to transform the rates and work with a linear model than to graduate the crude rates using a more mathematically demanding nonlinear model. 
The same philosophy applies in nonparametric graduation. 

Although several transformations $t$ exist (see, e.g., \citealp{Carr:Rupp:Tran:1988}, \citealp{Cox:Snel:Anal:1989}, and \citealp{Elan:John:Surv:1980}), the most commonly used in binary analysis is the logit (or log-odds) transformation
\begin{equation}
\mathring{q}_x^t=\ln\frac{\mathring{q}_x}{1-\mathring{q}_x},\quad x\in\mathcal{X},
\label{eq:logit transformation}
\end{equation}
with back-transform, with respect to the more general model \eqref{eq:adaptive NW kernel},
\begin{displaymath}
\widehat{q}_x=\frac{\exp\left\{\displaystyle\sum_{y\in\mathcal{X}}K_{h_x}\left(y;m=x\right)\mathring{q}_y^t\right\}}{1+\exp\left\{\displaystyle\sum_{y\in\mathcal{X}}K_{h_x}\left(y;m=x\right)\mathring{q}_y^t\right\}},\quad	x\in\mathcal{X}.
\end{displaymath}
By smoothing on a logistic scale and then back-transforming, we are guaranteed that $\widehat{q}_x\in\left[0,1\right]$. This transformation also reflects the fact that small changes when the mortality rate is near zero are as important as
larger changes when the mortality rate is much higher. 
\citet{Rens:Actu:1991} provides further motivation for this transformation, based on the theory of generalized
linear models.
The logit transformation \eqref{eq:logit transformation} is considered by the \code{dbkGrad} function via the argument specification \code{logit=T}.
However, because the choice of a transformation remains subjective, and the relative success of a particular transformation seems to depend on the data set \citep{GHVe:grad:1995}, the default setting of the \code{dbkGrad} function is \code{logit=F}.

\subsection[Pointwise confidence intervals]{Pointwise confidence intervals}
\label{subsec:pointwise confidence intervals}

In visual inspection and graphical interpretation of the estimated kernel sequence of points, pointwise confidence intervals at the considered ages $x\in\mathcal{X}$ provide relevant information, because they indicate the extent to which the estimates are well defined on $\mathcal{X}$.
Moreover, they are useful when nonparametric and parametric models are compared.
In the following formulas of this section, the bandwidth $h$, and eventually the sensitivity parameter $s$, are considered as \textit{a priori} fixed/selected. 

Since $\widehat{q}_x$ is a linear function of the mortality rates, as can be easily seen from \eqref{eq:NW kernel} and \eqref{eq:adaptive NW kernel}, and being $d_x\sim\text{Bin}\left(e_x,q_x\right)$
\begin{eqnarray*}
\VAR\left(\widehat{q}_x\right)
&=&\displaystyle\sum_{y\in\mathcal{X}}\left[K_{h_x}\left(y;m=x\right)\right]^2 \VAR\left(q_y\right)\\
&=&\displaystyle\sum_{y\in\mathcal{X}}\left[K_{h_x}\left(y;m=x\right)\right]^2 \VAR\left(\frac{d_y}{e_y}\right)\\
&=&\displaystyle\sum_{y\in\mathcal{X}}\left[K_{h_x}\left(y;m=x\right)\right]^2 \frac{q_y\left(1-q_y\right)}{e_y}. 
\end{eqnarray*}      
The above formula holds if independence of the $d_y$s is assumed and requires the knowledge of the number $e_y$ of exposed to risk at each age.
Substituting $q_y$ for $\widehat{q}_y$ yields the $\left(1-\alpha\right)\cdot 100\%$ pointwise confidence intervals 
\begin{equation}
\widehat{q}_x\mp z_{1-\frac{\alpha}{2}}\sqrt{\displaystyle\sum_{y\in\mathcal{X}}\left[K_{h_x}\left(y;m=x\right)\right]^2 \frac{\widehat{q}_y\left(1-\widehat{q}_y\right)}{e_y}}, \label{eq:confidence bands for ICRF}
\end{equation} 
where $z_{1-\frac{\alpha}{2}}$ is such that $\Phi\left[z_{1-\frac{\alpha}{2}}\right]=1-\frac{\alpha}{2}$.

\section[Package DBKGrad in use]{Package \pkg{DBKGrad} in use: the Sicily2008M data}
\label{sec:Package DBKGrad in use}

This tutorial uses the Sicily2008M dataset included in the \pkg{DBKGrad} package (also downloadable from \url{http://demo.istat.it/}) and already analyzed in \cite{Mazz:Punz:Grad:2012}.
Data consist of values for $\mathring{q}_x$ and $e_x$, $x=0,1,\ldots,100$, and are relative to the male population of the Sicily Region (Italy) for the year 2008.

To begin the analysis, data are loaded in the following way
\begin{CodeInput}
R> data("Sicily2008M")
R> obsqx <- Sicily2008M$qx
R> ex <- Sicily2008M$ex
\end{CodeInput}
The last two commands are only specified to simplify the subsequent notation.
For a quick look at data, the following commands can be used
\begin{CodeChunk}
\begin{CodeInput}
R> head(Sicily2008M)
\end{CodeInput}
\begin{CodeOutput}
          qx    ex
0 0.00465217 24816
1 0.00026728 25774
2 0.00017643 25950
3 0.00012708 26422
4 0.00010655 26172
5 0.00011917 25976
\end{CodeOutput}
\end{CodeChunk}
\begin{CodeChunk}
\begin{CodeInput}
R> tail(Sicily2008M)
\end{CodeInput}
\begin{CodeOutput}
           qx  ex
95  0.2597134 799
96  0.2631388 486
97  0.2648867 349
98  0.2694343 220
99  0.2845016 127
100 0.3169072 266
\end{CodeOutput}
\end{CodeChunk}
The second step consists in creating a \code{dbkGrad} object.
This step performs the discrete beta kernel graduation and prepares the object for analysis using the available plots.
This can be obtained, for example, by the following command
\begin{CodeChunk}
\begin{CodeInput}
R> resFX1 <- dbkGrad(obsqx=obsqx, omega=85)
\end{CodeInput}
\begin{CodeOutput}
It.    0, RSS =    3.50182, Par. =      0.002
It.    1, RSS =     1.6692, Par. =  0.00107794
It.    2, RSS =    1.53706, Par. =  0.000766075
It.    3, RSS =     1.4829, Par. =  0.000228109
It.    4, RSS =    1.48256, Par. =  0.000609218
It.    5, RSS =    1.45709, Par. =  0.000510961
It.    6, RSS =    1.45177, Par. =  0.000314446
It.    7, RSS =    1.44462, Par. =  0.000412703
It.    8, RSS =    1.44425, Par. =  0.000390018
It.    9, RSS =    1.44424, Par. =  0.00039553
It.   10, RSS =    1.44423, Par. =  0.000393552
It.   11, RSS =    1.44423, Par. =  0.000393552
\end{CodeOutput}
\end{CodeChunk}
Here, the (old) ages of interest are reduced from $\omega=100$ to $\omega=85$ via the specification \code{omega=85};
this allows to make the graphical inspection of the next plots easier.
The function \code{dbkGrad} produces, by default, fixed discrete beta kernel graduation in which the bandwidth is estimated by minimizing the cross-validation statistic \eqref{eq:CV statistic} with the residuals given in \eqref{eq:proportional differences}.
The iterations from the cross-validation procedure are printed at video. 
Also by default, no preliminary transformation of the data is considered.

\begin{figure}[!ht]
\centering
\resizebox{0.87\textwidth}{!}{\includegraphics{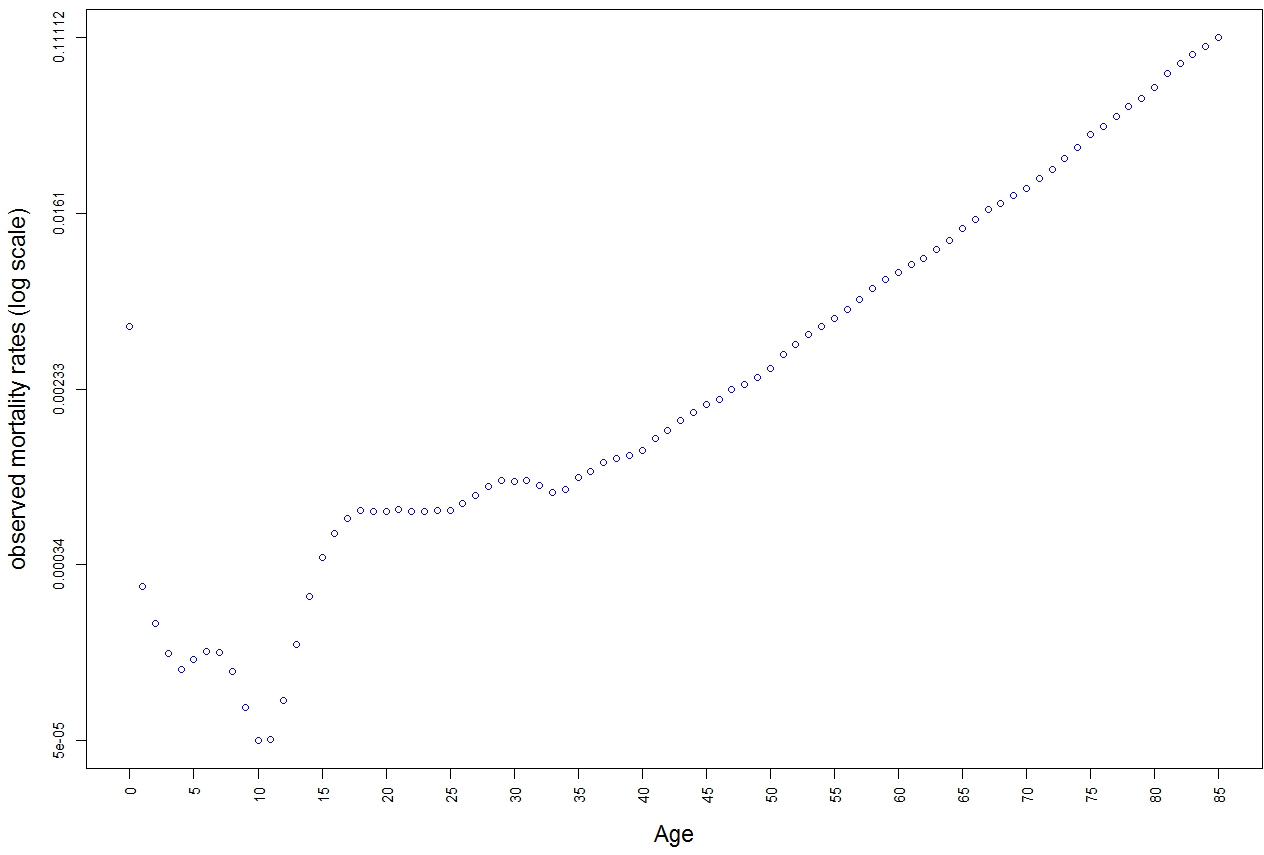}}
\caption{
Observed male mortality rates, in logarithmic scale, for the year 2008 in the Sicily Region (Italy).
\label{fig:FX_CV_obs} 
}
\end{figure}

\begin{figure}[!ht]
\centering
\resizebox{0.87\textwidth}{!}{\includegraphics{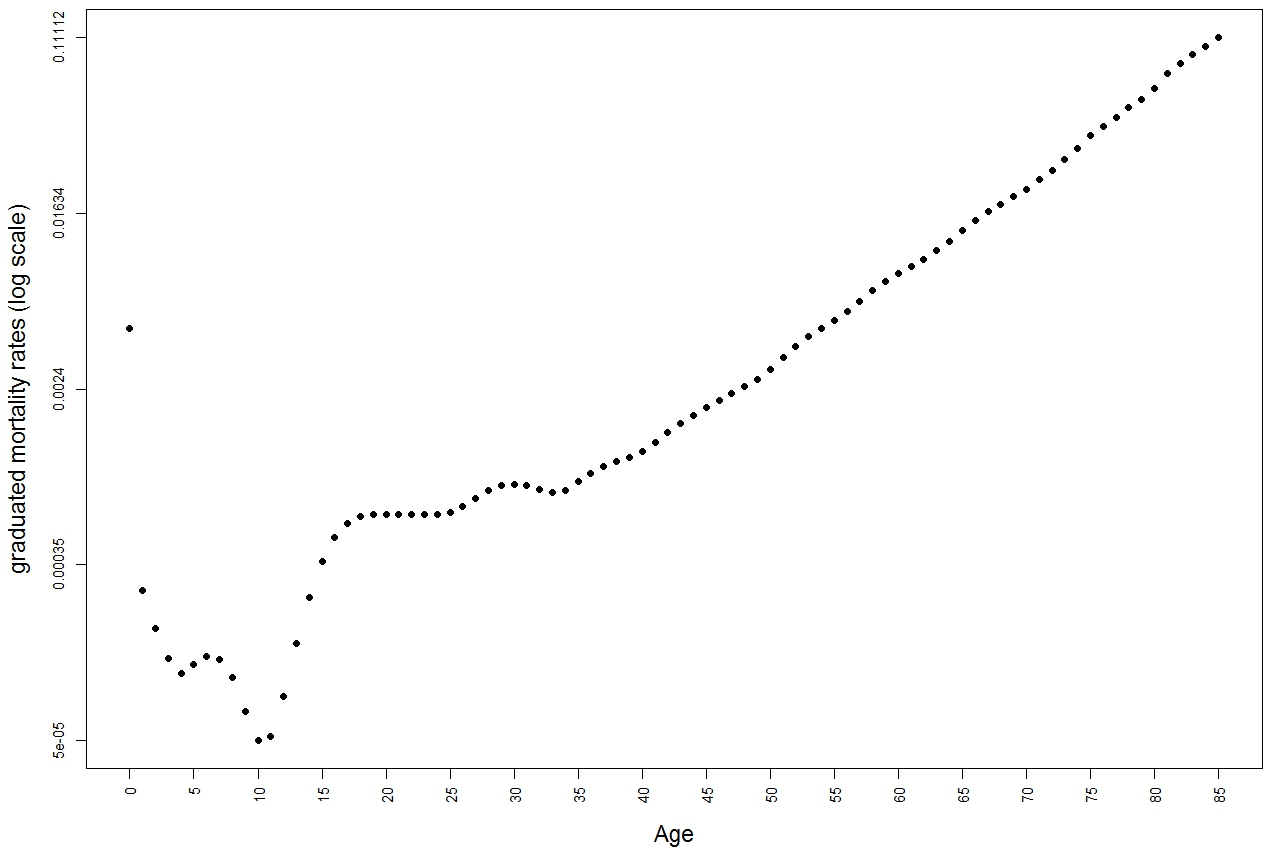}}
\caption{
Graduated male mortality rates, in logarithmic scale, for the year 2008 in the Sicily Region (Italy).
Graduation is made by the fixed discrete beta kernel estimator in \eqref{eq:NW kernel} where the bandwidth is estimated by minimizing the cross-validation statistic \eqref{eq:CV statistic} with residuals defined by \eqref{eq:proportional differences}. 
\label{fig:FX_CV_hat} 
}
\end{figure}

\begin{figure}[!ht]
\centering
\resizebox{0.87\textwidth}{!}{\includegraphics{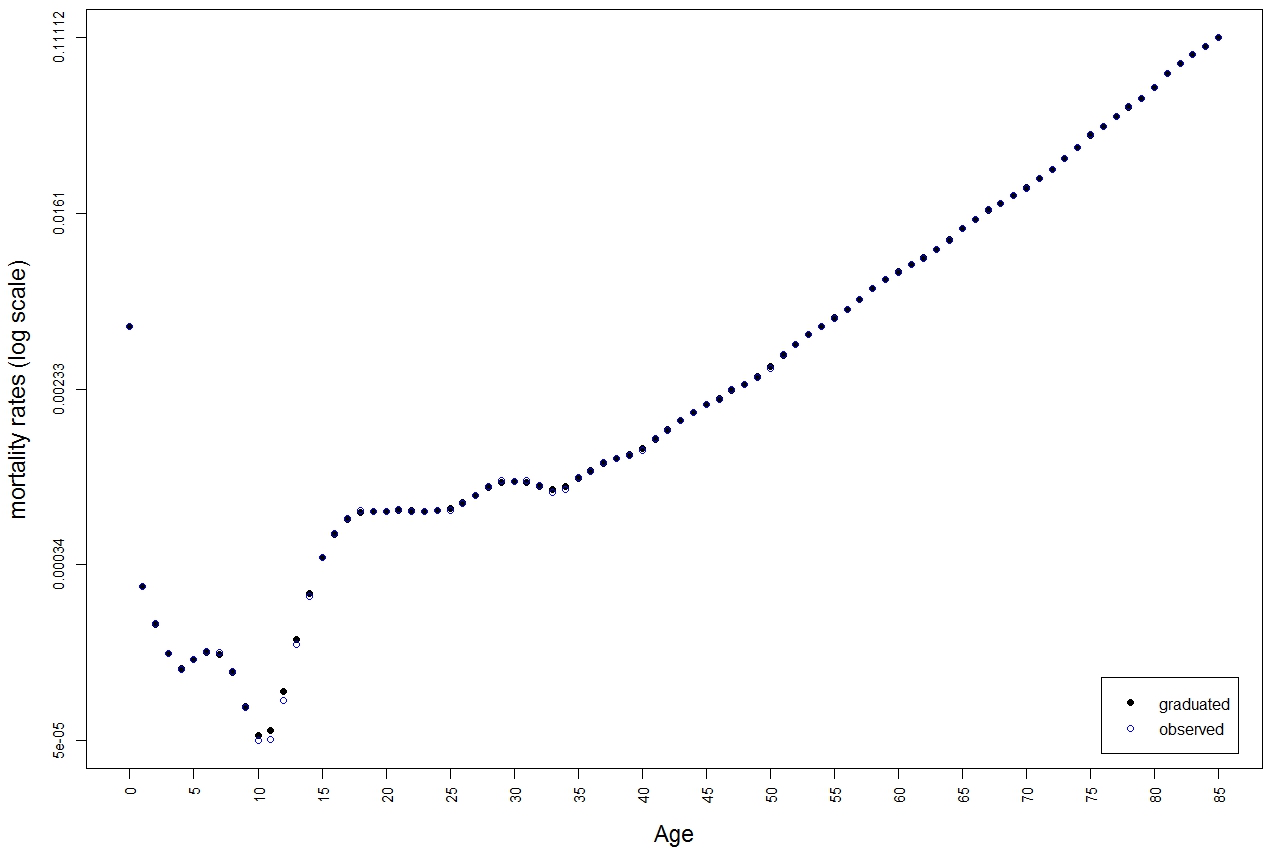}}
\caption{
Observed and graduated male mortality rates, in logarithmic scale, for the year 2008 in the Sicily Region (Italy).
Graduation is made by the fixed discrete beta kernel estimator in \eqref{eq:NW kernel} where the bandwidth is estimated by minimizing the cross-validation statistic \eqref{eq:CV statistic} with residuals defined by \eqref{eq:proportional differences}. 
\label{fig:FX_CV_hat_obs} 
}
\end{figure}

\begin{figure}[!ht]
\centering
\resizebox{0.80\textwidth}{!}{\includegraphics{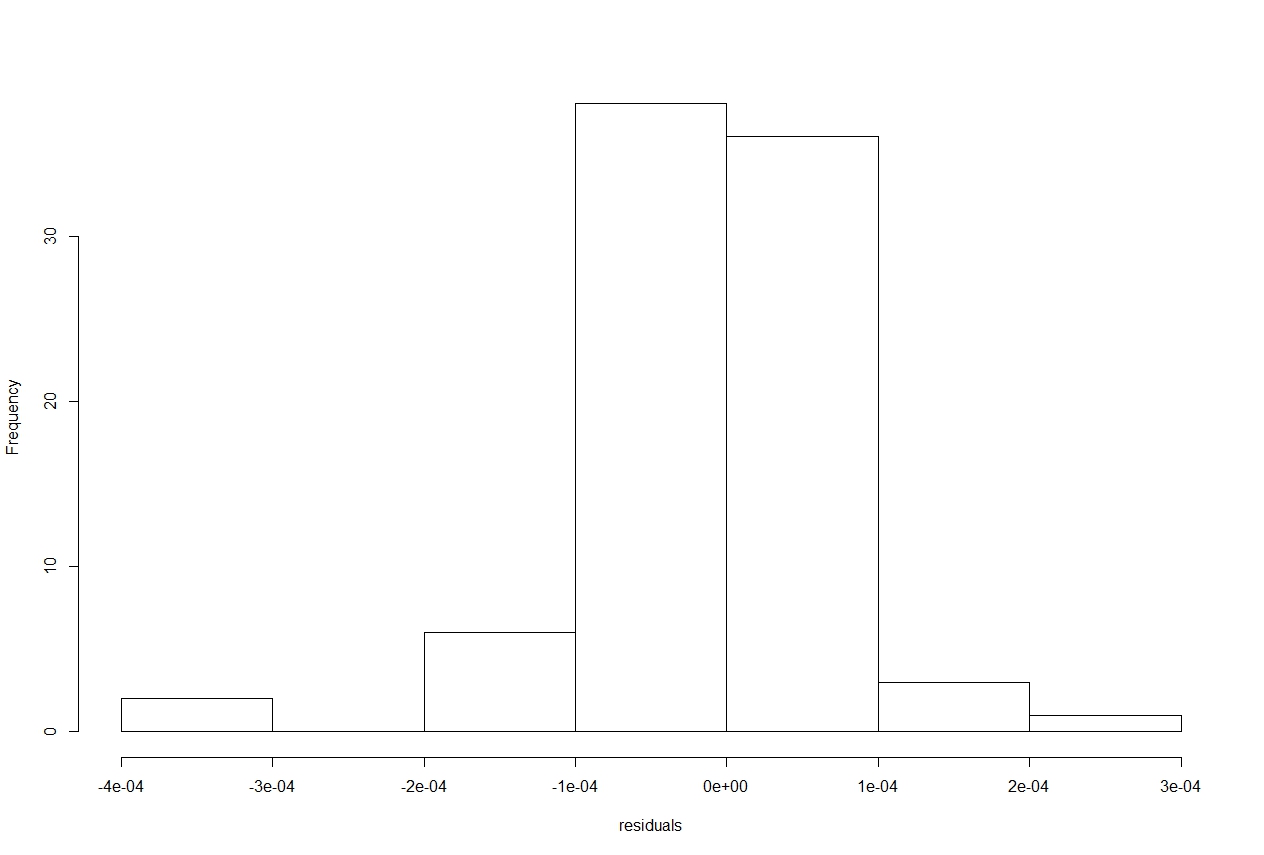}}
\caption{
Histogram of the residuals \eqref{eq:differences} arising from the fitting of the fixed discrete beta kernel estimator in \eqref{eq:NW kernel} to the mortality rates for the year 2008 in the Sicily Region (Italy). 
The bandwidth is estimated by minimizing the cross-validation statistic \eqref{eq:CV statistic} with residuals defined by \eqref{eq:proportional differences}. 
\label{fig:residuals} 
}
\end{figure}

Once the \code{dbkGrad} object \code{resFX1} is created, plots become available. 
The \code{plot} function allows for six different plots, that can be chosen by altering the \code{plottype} option.
The code 
\begin{CodeInput}
R> plot(resFX1, plottype="observed")
\end{CodeInput}
produces the plot of the crude mortality rates (\code{plottype="observed"}) in \figurename~\ref{fig:FX_CV_obs}, while the code 
\begin{CodeInput}
R> plot(resFX1, plottype="fitted")
\end{CodeInput}
produces the plot of the graduated mortality rates (\code{plottype="fitted"}) in \figurename~\ref{fig:FX_CV_hat}.
As usual in the graduation literature, a logarithmic scale is used.
In both the plots, a small but prominent hump, peaking around 18 years of age, is also visible; this ``excess mortality rate'', known in literature as accidental hump, is typically observed especially in males and it is probably due to an increase in a variety of risky activities, the most notable being to obtain a driver's license.
The simultaneous graphical representation of both crude and graduated mortality rates (see \figurename~\ref{fig:FX_CV_hat_obs}) is obtained via the command
\begin{CodeInput}
R> plot(resFX1, plottype="obsfit")
\end{CodeInput}
The histogram of the residuals \eqref{eq:differences}, displayed in \figurename~\ref{fig:residuals}, is obtained by the code
\begin{CodeInput}
R> plot(resFX1, plottype="histres")
\end{CodeInput}
It could be useful in model diagnostic checking.
The histogram of the proportional residuals \eqref{eq:proportional differences} can be obtained by specifying \code{plottype="histpropores"}.

To improve the graphical inspection of the obtained results, pointwise confidence interval can be added to the plot.
However, as said in Section~\ref{subsec:pointwise confidence intervals}, these intervals require the knowledge of the exposed to risk.
Thus, the \code{dbkGrad} object needs to be re-created to account for this aspect.
The code   
\begin{CodeInput}
R> resFX2 <- dbkGrad(obsqx, ex=ex, omega=85, alpha=0.05)
R> plot(resFX2, plottype="obsfit", CI=T)
\end{CodeInput}
produces the plot in \figurename~\ref{fig:FX_CV_hat_obs}.
\begin{figure}[!ht]
\centering
\resizebox{0.87\textwidth}{!}{\includegraphics{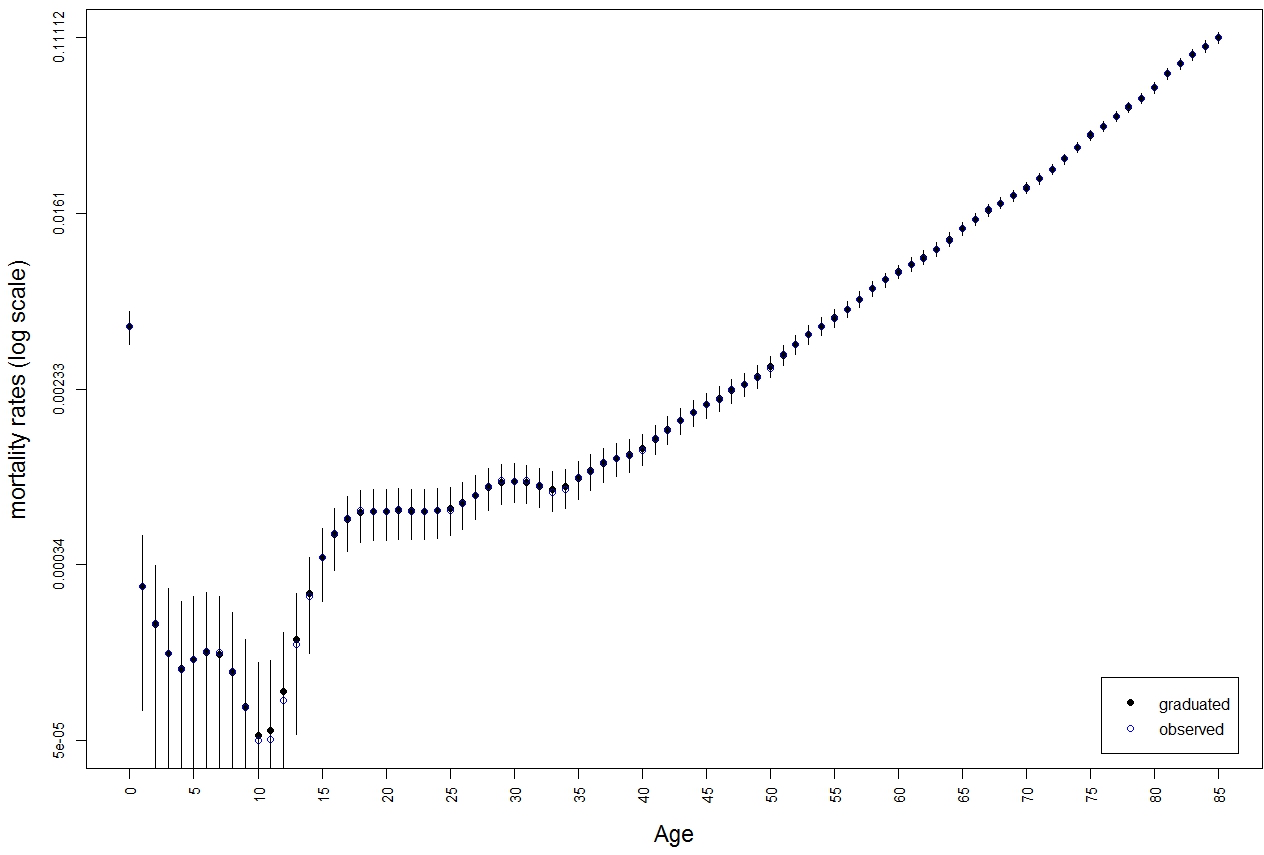}}
\caption{
Observed ans graduated male mortality rates, in logarithmic scale, for the year 2008 in the Sicily Region (Italy).
Graduation is made by the fixed discrete beta kernel estimator in \eqref{eq:NW kernel} where the bandwidth is estimated by minimizing the cross-validation statistic \eqref{eq:CV statistic} with residuals defined by \eqref{eq:proportional differences}.
Pointwise 95\% confidence intervals, computed as in \eqref{eq:confidence bands for ICRF}, are also superimposed. 
\label{fig:FX_CV_IC_hat_obs} 
}
\end{figure}
By the argument \code{ex=ex}, the exposed to risk are passed to the \code{dbkGrad} function. 
Also in the first row of code, the argument \code{alpha=0.05} -- which is the default -- specifies the value of $\alpha$ for the pointwise confidence intervals given in \eqref{eq:confidence bands for ICRF}.
In the plot command, the argument \code{plottype="fitted"} allows to display only the graduated sequence of points, while \code{CI=T} activates the pointwise confidence intervals, with the confidence level specified in the main function above.

\begin{figure}[!ht]
\centering
\resizebox{0.87\textwidth}{!}{\includegraphics{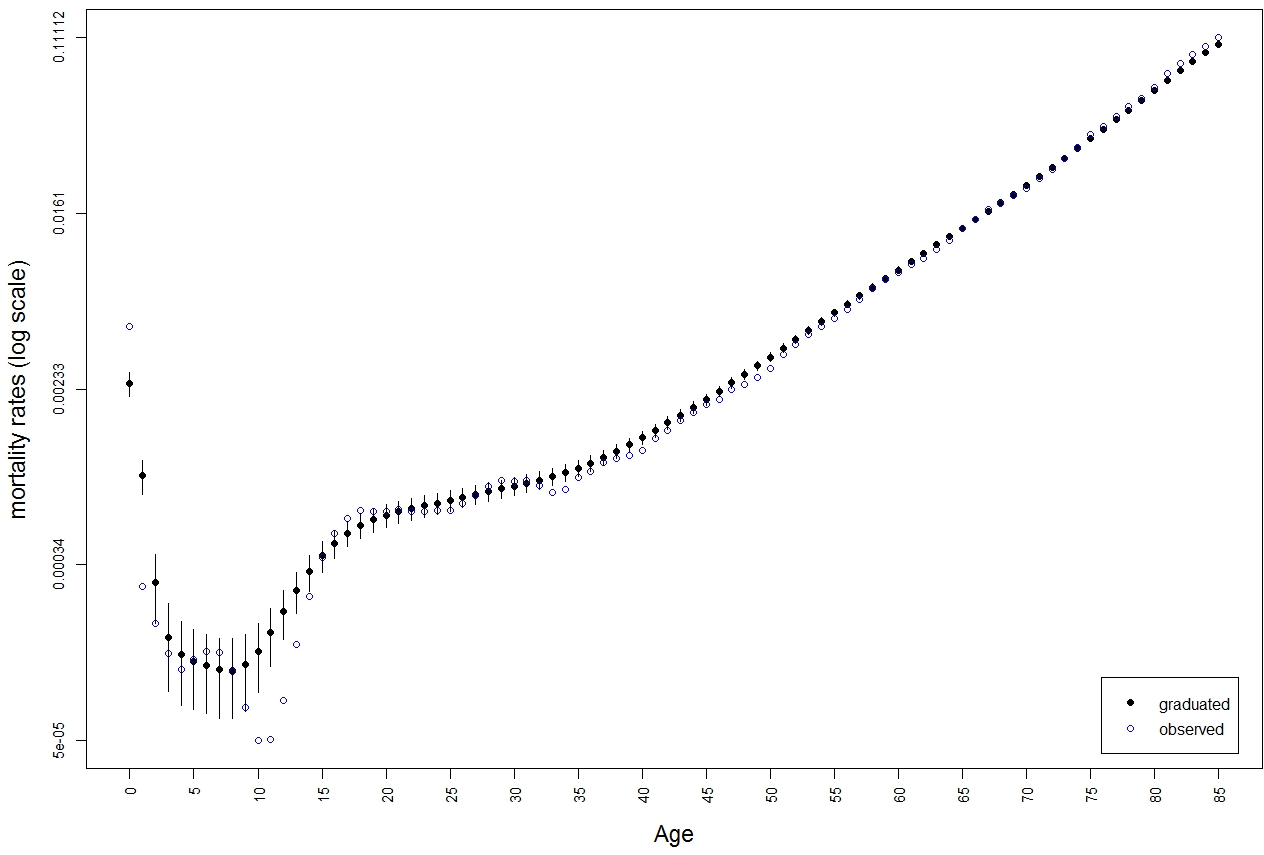}}
\caption{
Graduated male mortality rates, in logarithmic scale, for the year 2008 in the Sicily Region (Italy).
Graduation is made by the fixed discrete beta kernel estimator in \eqref{eq:NW kernel} where the bandwidth is manually specified.
Pointwise 95\% confidence intervals are also superimposed. 
\label{fig:FX_manual_IC_hat_obs} 
}
\end{figure}

Naturally, the user can specify a value for $h$ if, for example, he prefers an higher smoothness.
The code    
\begin{CodeInput}
R> resFX3 <- dbkGrad(obsqx, ex=ex, omega=85, h=0.01, cvh=F, alpha=0.05)
R> plot(resFX3, plottype="obsfit", CI=T)
\end{CodeInput}
produces the plot in \figurename~\ref{fig:FX_manual_IC_hat_obs} in which, if compared with \figurename~\ref{fig:FX_CV_hat_obs}, an higher smothness of the graduated sequence of points can be noted.
This is made possible by the (manual) specification \code{h=0.01}, along with the constraint \code{cvh=F} which avoids the cross-validation selection of the bandwidth.   

\begin{figure}[!ht]
\centering
\resizebox{0.80\textwidth}{!}{\includegraphics{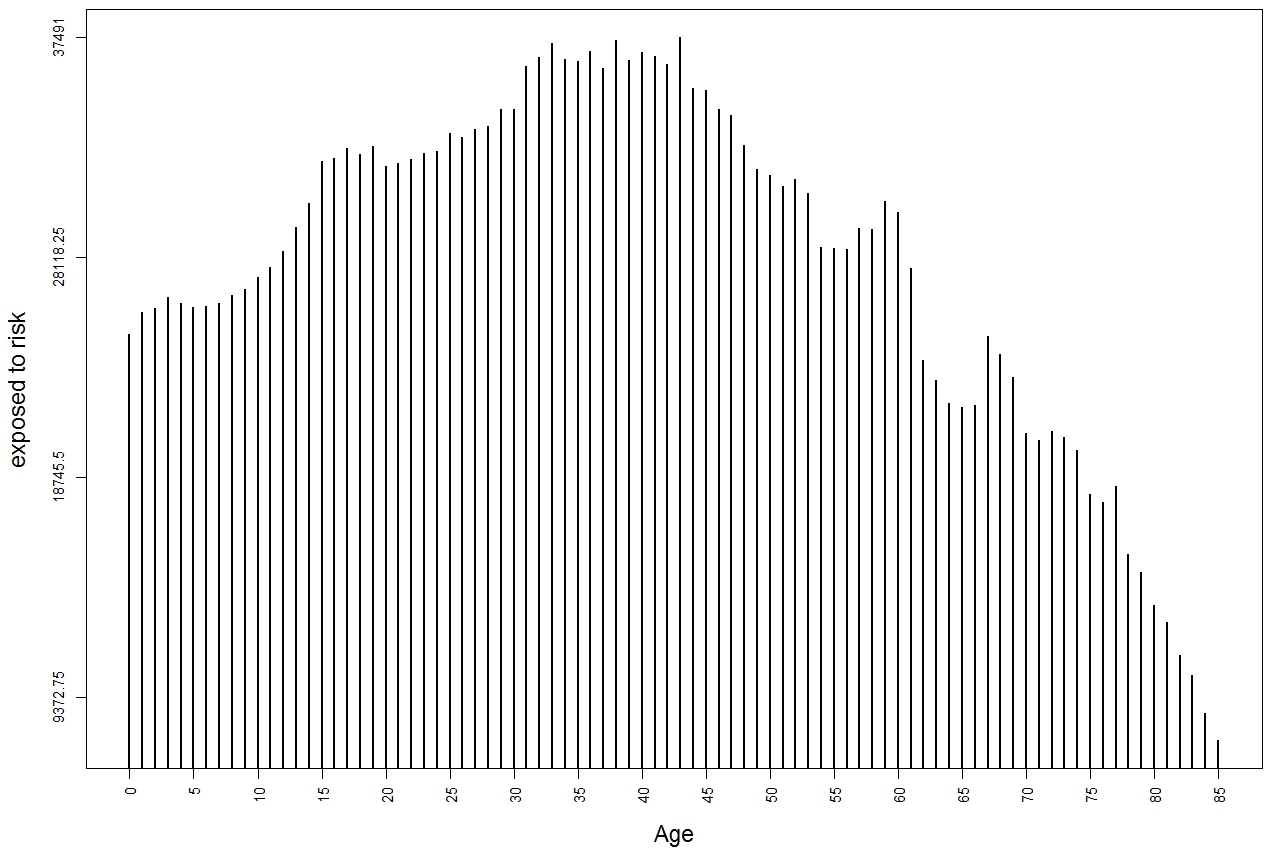}}
\caption{
Bar plot of the male exposure for the year 2008 in the Sicily Region. 
\label{fig:exposed} 
}
\end{figure}

\begin{figure}[!ht]
\centering
\resizebox{0.87\textwidth}{!}{\includegraphics{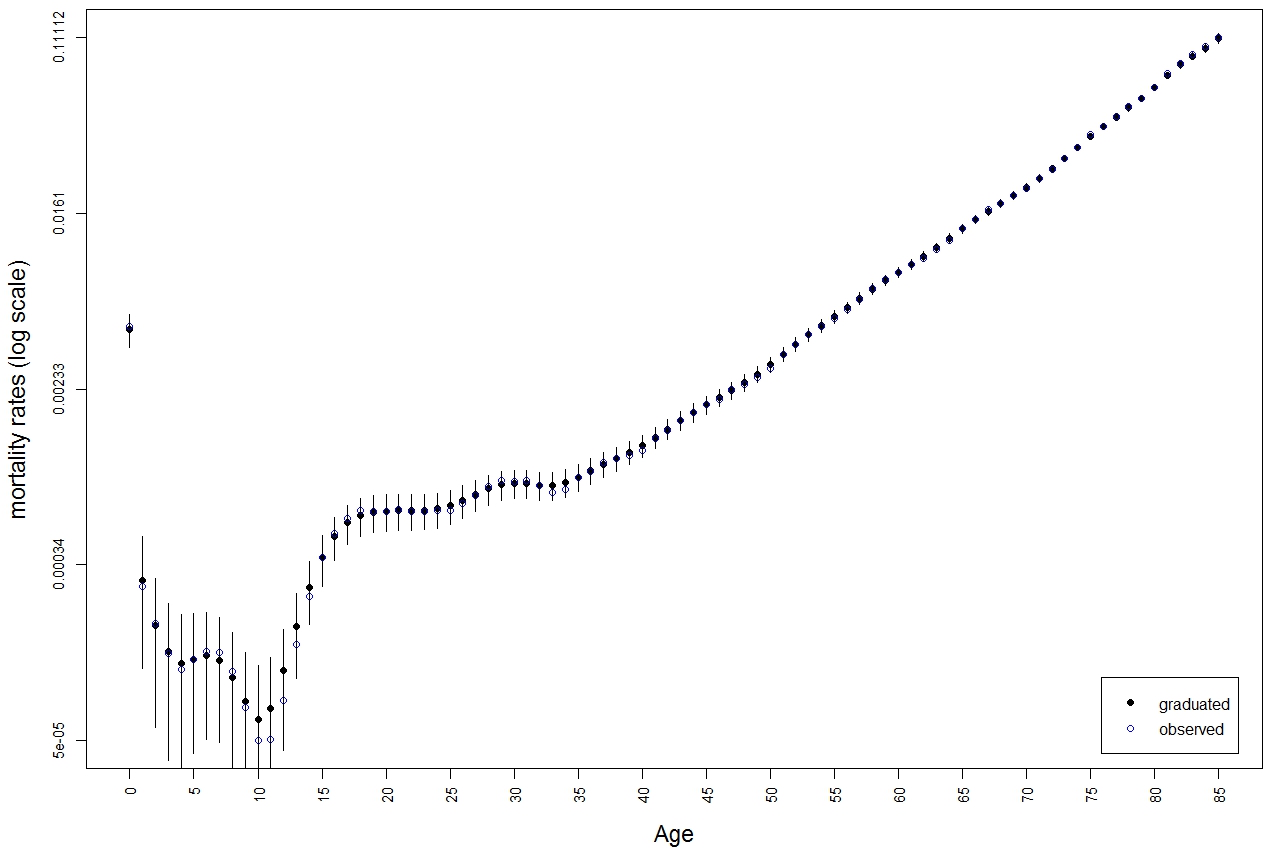}}
\caption{
Observed ans graduated male mortality rates, in logarithmic scale, for the year 2008 in the Sicily Region (Italy).
Graduation is made by the adaptive discrete beta kernel estimator in \eqref{eq:adaptive NW kernel} where the bandwidth is estimated by minimizing the (conditional) cross-validation statistic $\text{CV}\left(h|s=0.28\right)$ with residuals defined by \eqref{eq:differences}.
Pointwise 95\% confidence intervals, computed as in \eqref{eq:confidence bands for ICRF}, are also superimposed. 
\label{fig:EX_CVcond_IC_hat_obs} 
}
\end{figure}

So far, the exposed to risk have been only used to define the pointwise confidence intervals.
However, as explained in Section~\ref{sec:Making the bandwidth adaptive}, they are also useful to take into account the reliability of the data.
The code
\begin{CodeInput}
R> plot(resFX3, plottype="exposed")
\end{CodeInput}
produces the bar plot of the male population at risk displayed in \figurename~\ref{fig:exposed}.
The great variation in exposure, over the age range, shows the usefulness of an adaptive approach. 
Note that one offhanded change in exposure is visible in the age ranges 60--62, due to the Second World War.
The code
\begin{CodeChunk}
\begin{CodeInput}
R> resEX <- dbkGrad(obsqx, omega=85, ex, bandwidth="EX", s=0.28,  
 +                  cvres="res", cvh=T, cvs=F, alpha=0.05)
\end{CodeInput}
\begin{CodeOutput}
It.    0, RSS = 0.000223193, Par. =      0.002
It.    1, RSS = 0.000222536, Par. =  0.00227673
It.    2, RSS = 0.000222489, Par. =  0.00222072
It.    3, RSS = 0.000222489, Par. =  0.00221859
It.    4, RSS = 0.000222489, Par. =  0.00221859
\end{CodeOutput}
\end{CodeChunk}
reproduces the scheme followed in \citet{Mazz:Punz:Grad:2012} where, the sensitivity parameter is fixed to $s = 0.28$, the bandwidth $h$ is selected by minimizing the (conditional) cross-validation statistic $\text{CV}\left(h|s=0.28\right)$ in which the classical residuals in \eqref{eq:differences} are used via the specification \code{cvres="res"}.
The corresponding graphical representation in \figurename~\ref{fig:EX_CVcond_IC_hat_obs}
is obtained via the code
\begin{CodeInput}
R> plot(resEX, plottype="obsfit", CI=T)
\end{CodeInput}
It is easy to note that the estimated points have a more ``graduated'' behavior, with respect to the observed ones and to the graduated ones displayed in \figurename~\ref{fig:FX_CV_IC_hat_obs}, above all for the ages from 0 to 15.

\begin{figure}[!ht]
\centering
\resizebox{0.87\textwidth}{!}{\includegraphics{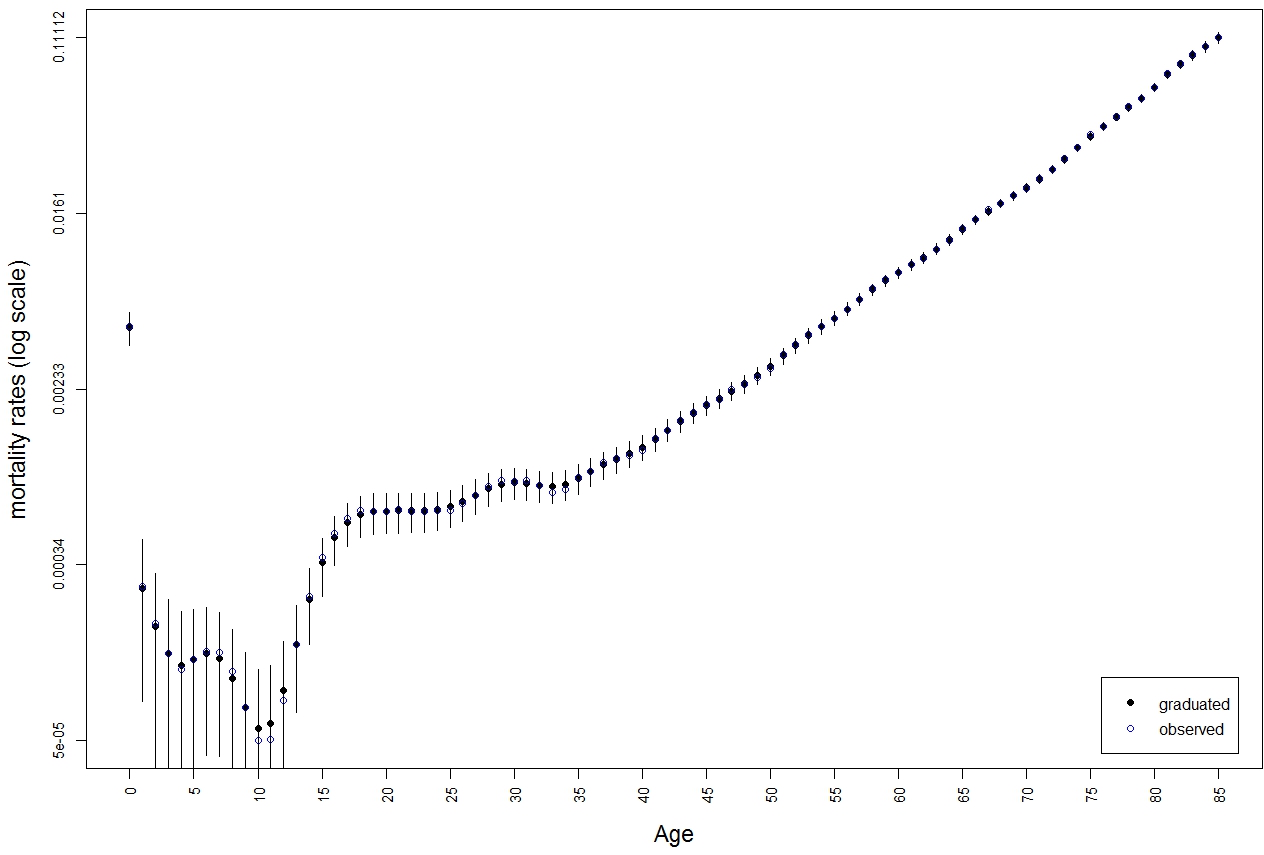}}
\caption{
Observed ans graduated male mortality rates, in logarithmic scale, for the year 2008 in the Sicily Region (Italy).
Graduation is made by the adaptive discrete beta kernel estimator in \eqref{eq:adaptive NW kernel} where the bandwidth is estimated by minimizing the (joint) cross-validation statistic $\text{CV}\left(h,s\right)$ with residuals defined by \eqref{eq:proportional differences}.
A preliminary logit transformation is applied to the rates.
Pointwise 95\% confidence intervals, computed as in \eqref{eq:confidence bands for ICRF}, are also superimposed. 
\label{fig:VC_CVjoint_IC_hat_obs} 
}
\end{figure}

The code 
\begin{CodeChunk}
\begin{CodeInput}
R> resVC <- dbkGrad(obsqx, omega=85, ex, logit=T, bandwidth="VC", 
 +                  cvh=T, cvs=T, alpha=0.05)
R> plot(resVC, plottype="obsfit", CI=T)
\end{CodeInput}
\begin{CodeOutput}
It.    0, RSS =   0.299308, Par. =      0.002        0.2
It.    1, RSS =    0.29835, Par. =  0.00181587  0.0934201
It.    2, RSS =    0.29803, Par. =  0.0013289  0.0340463
It.    3, RSS =   0.297901, Par. =  0.00114724  0.00665506
It.    4, RSS =   0.297849, Par. =  0.00101142          0
It.    5, RSS =   0.297849, Par. =  0.00101145          0
It.    6, RSS =   0.297849, Par. =  0.00101147          0
It.    7, RSS =   0.297849, Par. =  0.00101148          0
It.    8, RSS =   0.297849, Par. =  0.00101149          0
It.    9, RSS =   0.297849, Par. =  0.00101149          0
It.   10, RSS =   0.297849, Par. =  0.0010115          0
It.   11, RSS =   0.297849, Par. =  0.0010115          0
It.   12, RSS =   0.297849, Par. =  0.0010115          0
It.   13, RSS =   0.297849, Par. =  0.0010115          0
It.   14, RSS =   0.297849, Par. =  0.0010115          0
\end{CodeOutput}
\end{CodeChunk}
allows to further show the package flexibility.
The graphical result is displayed in \figurename~\ref{fig:VC_CVjoint_IC_hat_obs}.
In particular, a preliminary logit transformation of the mortality rates is applied (\code{logit=T}), as explained in Section~\ref{subsec:Transforming mortality rates} and, via the specifications \code{cvh=T} and \code{cvs=T}, both $h$ and $s$ are automatically selected by minimizing the joint cross-validation score $\text{CV}\left(h,s\right)$. 
An effect of having applied the logit transformation is that, since this time cross-validation selects $s=0$, there is no need of using the adaptive variant of the discrete beta kernel estimator.

Finally, the code 
\begin{CodeChunk}
\begin{CodeInput}
R> as.data.frame(resVC)
\end{CodeInput}
\begin{CodeOutput}
        obsqx fitted.values exposed   lowerbound   upperbound
0  0.00465217  4.583428e-03   24816 3.747432e-03 0.0054194238
1  0.00026728  2.610820e-04   25774 7.551190e-05 0.0004466521
2  0.00017643  1.714462e-04   25950 3.640966e-05 0.0003064828
3  0.00012708  1.269107e-04   26422 2.372682e-05 0.0002300946
4  0.00010655  1.118216e-04   26172 2.192182e-05 0.0002017214
5  0.00011917  1.193277e-04   25976 3.197348e-05 0.0002066819
.           .             .       .            .            .
.           .             .       .            .            .
.           .             .       .            .            .
80 0.06426613  6.410928e-02   13268 6.138194e-02 0.0668366143
81 0.07523889  7.380717e-02   12551 7.062169e-02 0.0769926510
82 0.08380302  8.287889e-02   11144 7.900452e-02 0.0867532688
83 0.09219433  9.134057e-02   10276 8.661340e-02 0.0960677363
84 0.10101198  1.004669e-01    8676 9.451396e-02 0.1064199097
85 0.11111700  1.110619e-01    7525 1.039997e-01 0.1181241766
\end{CodeOutput}
\end{CodeChunk}
summarizes to the user, via a dataframe, the table of all the most important quantities for each age.
This may be useful when exporting the results. 

%
%

\section[Conclusions]{Conclusions}
\label{sec:conclusions}

In this paper we have presented the \pkg{DBKGrad} package for the \proglang{R} environment. 
This package is specifically conceived for nonparametric graduation of discrete finite functions, such are mortality rates. 
The package is conceptually simple and easy to use; nevertheless several options are available to the user. 
He may choose among fixed and adaptive bandwidths, the latter being based, via two different formulations, on the exposed to the risk of dying. 
Furthermore, the bandwidth and/or a dampening factor may be indicated by the user or chosen by cross-validation; the cross-validation score being minimized may be based on the traditional sum of squared residuals or on an alternative formulation used in the graduation literature, that is the squared proportional residuals. 
Several plots of either types of residuals, as well as of observed data and of fitted data with confidence intervals, are provided. 
The package also included an illustrative data set, which contains mortality data for the 2008 male population in the Region of Sicily (Italy).
We believe that the \pkg{DBKGrad} package may prove useful to actuaries, demographers,  and other social scientists, either as a modeling tool or, if parametric models are to be used, it may still be useful for carrying  out a diagnosis of parametric models or simply to examine data.

\setlength{\bibsep}{3.5pt} 
\bibliography{References}

\begin{thebibliography}{26}
\newcommand{\enquote}[1]{``#1''}
\providecommand{\natexlab}[1]{#1}
\providecommand{\url}[1]{\texttt{#1}}
\providecommand{\urlprefix}{URL }
\expandafter\ifx\csname urlstyle\endcsname\relax
  \providecommand{\doi}[1]{doi:\discretionary{}{}{}#1}\else
  \providecommand{\doi}{doi:\discretionary{}{}{}\begingroup
  \urlstyle{rm}\Url}\fi
\providecommand{\eprint}[2][]{\url{#2}}

\bibitem[{Bagnato and Punzo(in press)}]{Bagn:Punz:Fine:2013}
Bagnato L, Punzo A (in press).
\newblock \enquote{{Finite Mixtures of Unimodal Beta and Gamma Densities and
  the $k$-Bumps Algorithm}.}
\newblock \emph{Computational Statistics}.

\bibitem[{Carroll and Ruppert(1988)}]{Carr:Rupp:Tran:1988}
Carroll RJ, Ruppert D (1988).
\newblock \emph{{Transformation and Weighting in Regression}}, volume~30.
\newblock Chapman \& Hall, New York.

\bibitem[{Chen(2000)}]{Chen:beta:2000}
Chen S (2000).
\newblock \enquote{Beta Kernel Smoothers for Regression Curves.}
\newblock \emph{Statistica Sinica}, \textbf{10}(1), 73--91.

\bibitem[{Copas and Haberman(1983)}]{Copa:nonp:1983}
Copas JB, Haberman S (1983).
\newblock \enquote{{Non-parametric Graduation Using Kernel Methods}.}
\newblock \emph{Journal of the Institute of Actuaries}, \textbf{110}(1),
  135--156.

\bibitem[{Cox and Snell(1989)}]{Cox:Snel:Anal:1989}
Cox DR, Snell E (1989).
\newblock \emph{{Analysis of Binary Data}}.
\newblock Chapman \& Hall, New York.

\bibitem[{Debòn \emph{et~al.}(2006)Debòn, Montes, and Sala}]{DMSa:acom:2006}
Debòn A, Montes F, Sala R (2006).
\newblock \enquote{{A Comparison of Nonparametric Methods in the Graduation of
  Mortality: Application to Data from the Valencia Region (Spain)}.}
\newblock \emph{International Statistical Review}, \textbf{74}(2), 215--233.

\bibitem[{Elandt-Johnson and Johnson(1980)}]{Elan:John:Surv:1980}
Elandt-Johnson R, Johnson N (1980).
\newblock \emph{{Survival Models and Data Analysis}}.
\newblock John Wiley \& Sons, Canada.

\bibitem[{Elzhov \emph{et~al.}(2010)Elzhov, Mullen, and Bolker}]{minpack.lm}
Elzhov TV, Mullen KM, Bolker B (2010).
\newblock \emph{\pkg{minpack.lm}: \proglang{R} Interface to the
  Levenberg-Marquardt Nonlinear Least-Squares Algorithm Found in
  \proglang{MINPACK}}.
\newblock \proglang{R} package version 1.1-5.
\newblock \urlprefix\url{http://CRAN.R-project.org/package=minpack.lm}.

\bibitem[{Gavin \emph{et~al.}(1993)Gavin, Haberman, and
  Verrall}]{Gavi:Habe:Verr:Movi:1993}
Gavin J, Haberman S, Verrall R (1993).
\newblock \enquote{{Moving Weighted Average Graduation Using Kernel
  Estimation}.}
\newblock \emph{Insurance: Mathematics and Economics}, \textbf{12}(2),
  113--126.

\bibitem[{Gavin \emph{et~al.}(1994)Gavin, Haberman, and
  Verrall}]{GHVe:onth:1994}
Gavin J, Haberman S, Verrall R (1994).
\newblock \enquote{{On the Choice of Bandwidth for Kernel Graduation}.}
\newblock \emph{Journal of the Institute of Actuaries}, \textbf{121}(1),
  119--134.

\bibitem[{Gavin \emph{et~al.}(1995)Gavin, Haberman, and
  Verrall}]{GHVe:grad:1995}
Gavin J, Haberman S, Verrall R (1995).
\newblock \enquote{{Graduation by Kernel and Adaptive Kernel Methods with a
  Boundary Correction}.}
\newblock \emph{Transactions of Society of Actuaries}, \textbf{47}, 173--209.

\bibitem[{Haberman and Renshaw(1996)}]{Habe:Rens:Gene:1996}
Haberman S, Renshaw A (1996).
\newblock \enquote{{Generalized Linear Models and Actuarial Science}.}
\newblock \emph{The Statistician}, \textbf{45}(4), 407--436.

\bibitem[{H{\"a}rdle(1992)}]{Hard:Appl:1992}
H{\"a}rdle W (1992).
\newblock \emph{{Applied Nonparametric Regression}}, volume~19 of
  \emph{Econometric Society Monographs}.
\newblock Cambridge University Press, Cambridge.

\bibitem[{Hayfield and Racine(2008)}]{Hayfield:Racine:2008:JSSOBK:v27i05}
Hayfield T, Racine JS (2008).
\newblock \enquote{Nonparametric Econometrics: The \pkg{np} Package.}
\newblock \emph{Journal of Statistical Software}, \textbf{27}(5), 1--32.
\newblock ISSN 1548-7660.
\newblock \urlprefix\url{http://www.jstatsoft.org/v27/i05}.

\bibitem[{Heligman and Pollard(1980)}]{Heli:Poll:Thea:1980}
Heligman L, Pollard J (1980).
\newblock \enquote{{The Age Pattern of Mortality}.}
\newblock \emph{Journal of the Institute of Actuaries}, \textbf{107}(1),
  49--80.

\bibitem[{London(1985)}]{Lond:Grad:1985}
London D (1985).
\newblock \emph{Graduation: The Revision of Estimates}.
\newblock Actex publications Abington, Connecticut.

\bibitem[{Mazza and Punzo(2011)}]{Mazz:Punz:Disc:2011}
Mazza A, Punzo A (2011).
\newblock \enquote{{Discrete Beta Kernel Graduation of Age-Specific Demographic
  Indicators}.}
\newblock In S~Ingrassia, R~Rocci, M~Vichi (eds.), \emph{New Perspectives in
  Statistical Modeling and Data Analysis}, volume~42 of \emph{Studies in
  Classification, Data Analysis and Knowledge Organization}, pp. 127--134.
  Springer-Verlag, Berlin-Heidelberg.

\bibitem[{Mazza and Punzo(2013)}]{Mazz:Punz:Grad:2012}
Mazza A, Punzo A (2013).
\newblock \enquote{{Graduation by Adaptive Discrete Beta Kernels}.}
\newblock In A~Giusti, G~Ritter, M~Vichi (eds.), \emph{Classification and Data
  Mining}, volume~44 of \emph{Studies in Classification, Data Analysis and
  Knowledge Organization}, pp. 77--84. Springer-Verlag, Berlin-Heidelberg.

\bibitem[{Mazza and Punzo(in press)}]{Mazz:Punz:Usin:2013}
Mazza A, Punzo A (in press).
\newblock \enquote{{Using the Variation Coefficient for Adaptive Discrete Beta
  Kernel Graduation}.}
\newblock In P~Giudici, S~Ingrassia, M~Vichi (eds.), \emph{Advances in
  Statistical Modelling for Data Analysis}, Studies in Classification, Data
  Analysis and Knowledge Organization. Springer-Verlag, Berlin-Heidelberg.

\bibitem[{Moré(1978)}]{More:TheL:1978}
Moré J (1978).
\newblock \enquote{{The Levenberg-Marquardt Algorithm: Implementation and
  Theory}.}
\newblock In G~Watson (ed.), \emph{Numerical Analysis}, volume 630 of
  \emph{Lecture Notes in Mathematics}, pp. 104--116. Springer-Verlag,
  Berlin-Heidelberg.

\bibitem[{Peristera and Kostaki(2005)}]{Peri:Kost:Anev:2005}
Peristera P, Kostaki A (2005).
\newblock \enquote{{An Evaluation of the Performance of Kernel Estimators for
  Graduating Mortality Data}.}
\newblock \emph{Journal of Population Research}, \textbf{22}(2), 185--197.

\bibitem[{Punzo(2010)}]{Punz:disc:2010}
Punzo A (2010).
\newblock \enquote{{Discrete Beta-type Models}.}
\newblock In H~Locarek-Junge, C~Weihs (eds.), \emph{Classification as a Tool
  for Research}, volume~40 of \emph{Studies in Classification, Data Analysis
  and Knowledge Organization}, pp. 253--261. Springer-Verlag,
  Berlin-Heidelberg.

\bibitem[{Punzo and Zini(2012)}]{Punz:Zini:Appr:2012}
Punzo A, Zini A (2012).
\newblock \enquote{{Discrete Approximations of Continuous and Mixed Measures on
  a Compact Interval}.}
\newblock \emph{Statistical Papers}, \textbf{53}(3), 563--575.

\bibitem[{{\proglang{R} Development Core Team}(2012)}]{R}
{\proglang{R} Development Core Team} (2012).
\newblock \emph{\proglang{R}: A Language and Environment for Statistical
  Computing}.
\newblock \proglang{R} Foundation for Statistical Computing, Vienna, Austria.
\newblock {ISBN} 3-900051-07-0, \urlprefix\url{http://www.R-project.org/}.

\bibitem[{Renshaw(1991)}]{Rens:Actu:1991}
Renshaw A (1991).
\newblock \enquote{{Actuarial Graduation Practice and Generalized Linear and
  Non-Linear Models}.}
\newblock \emph{Journal of the Institute of Actuaries}, \textbf{118}(2),
  295--312.

\bibitem[{Stone(1974)}]{Ston:cros:1974}
Stone M (1974).
\newblock \enquote{Cross-Validatory Choice and Assessment of Statistical
  Predictions.}
\newblock \emph{Journal of the Royal Statistical Society: Series B},
  \textbf{36}(1), 111--147.

\end{thebibliography}

\end{document}